\documentclass[%
 reprint,
 twocolumn
 superscriptaddress,
 showpacs,
 preprintnumbers,
 amsmath,amssymb,
 aps,
]{revtex4-1}

\DeclareMathAlphabet{\mathsfit}{T1}{\sfdefault}{\mddefault}{\sldefault}
\SetMathAlphabet{\mathsfit}{bold}{T1}{\sfdefault}{\bfdefault}{\sldefault}

\usepackage{graphicx}
\graphicspath{{fig/}}   
\usepackage{dcolumn}
\usepackage{bm}
\usepackage{makecell}
\usepackage{braket} 
\usepackage{siunitx}
\usepackage{color}
\usepackage{xr}
\usepackage{upgreek}
\usepackage[colorlinks]{hyperref}
\usepackage[capitalize]{cleveref}
\usepackage[caption=false]{subfig}
\usepackage{multirow} 
\usepackage{threeparttable,array} 
\usepackage{amssymb}

\DeclareMathSymbol{\shortminus}{\mathbin}{AMSa}{"39}

\makeatletter

\newcommand{\Rmnum}[1]{\expandafter\@slowromancap\romannumeral #1@}
\makeatother

\begin{document}

	
	\title{Accelerated quantum adiabatic transfer in superconducting qubits}
	
	\author{Wen Zheng}
	\thanks{These authors contributed equally to this work.}
	\affiliation{National Laboratory of Solid State Microstructures, School of Physics, Nanjing University, Nanjing 210093, China}
	\author{Jianwen Xu}
	\thanks{These authors contributed equally to this work.}
	\affiliation{National Laboratory of Solid State Microstructures, School of Physics, Nanjing University, Nanjing 210093, China}

	\author{Zhimin Wang}
	\affiliation{National Laboratory of Solid State Microstructures, School of Physics, Nanjing University, Nanjing 210093, China}
	\author{Yuqian Dong}
	\affiliation{National Laboratory of Solid State Microstructures, School of Physics, Nanjing University, Nanjing 210093, China}

	\author{Dong Lan}
	\affiliation{National Laboratory of Solid State Microstructures, School of Physics, Nanjing University, Nanjing 210093, China}
	\author{Xinsheng Tan}
	\email{tanxs@nju.edu.cn}
	\affiliation{National Laboratory of Solid State Microstructures, School of Physics, Nanjing University, Nanjing 210093, China}
	\author{Yang Yu}
	\email{yuyang@nju.edu.cn}
	\affiliation{National Laboratory of Solid State Microstructures, School of Physics, Nanjing University, Nanjing 210093, China}

	\date{\today}

\begin{abstract}    
	{
        Quantum adiabatic transfer is widely used in quantum computation and quantum simulation.
        However, the transfer speed is limited by the quantum adiabatic approximation condition, which hinders its application in quantum systems with a short decoherence time.
        Here we demonstrate quantum adiabatic state transfers that jump along geodesics in one-qubit and two-qubit superconducting transmons.
        This approach possesses the advantages of speed, robustness, and high fidelity compared with the usual adiabatic process.
        Our protocol provides feasible strategies for improving state manipulation and gate operation in superconducting quantum circuits.
        }    
\end{abstract}


\maketitle

\section{Introduction}
\label{sec:intro}

    Superconducting quantum circuits \cite{doi:10.1126/science.1231930, you_atomic_2011, arute2019quantum, gong2021quantum, PhysRevLett.127.180501} provide an excellent platform for quantum computation and quantum simulation.
	With the increase of qubit integration, further improvement in the fidelity of quantum state manipulation and gate operation is constrained by crosstalk, leakage, and relatively short decoherence times. 
	Quantum adiabatic evolution, which is robust against local fluctuations, has been proposed as an alternative approach for conducting universal quantum gate operations \cite{dicarlo2009demonstration, PhysRevA.90.022307,PhysRevA.65.012322, PhysRevLett.125.240503} and quantum simulation \cite{roushan_observation_2014}.
    Historically, the quantum adiabatic concept played an important role in the progress of fundamental physics including Landau-Zener transition (LZT) \cite{landau1932, zener1932non}, geometric phases \cite{berry1984quantal, PhysRevLett.58.1593, PhysRevLett.51.2167, shapere1989geometric}, and topology \cite{nakahara2018geometry}.
    Recently, it has been applied in quantum algorithm \cite{farhi2001quantum, PhysRevA.65.042308}, quantum sensing \cite{RevModPhys.89.035002}, quantum annealing \cite{PhysRevE.58.5355, doi:10.1126/science.284.5415.779, johnson2011quantum}, and quantum coherent control \cite{RevModPhys.89.015006, kumar2016stimulated, xu2016coherent, vepsalainen2019superadiabatic, yang2019realization, PhysRevLett.124.240502, li2021coherent}.
	Quantum adiabatic control has been used in many active research areas including quantum computation \cite{nielsen_chuang_2010, RevModPhys.90.015002, aharonov2008adiabatic} and quantum simulation \cite{feynman_simulating_nodate, RevModPhys.86.153, PRXQuantum.2.017003}.

    The essence of quantum adiabatic control is the quantum adiabatic theorem \cite{ehrenfest1916adiabatische, born1928beweis, PhysRev.51.648, kato1950adiabatic}, which states that the quantum state $|\Psi(t)\rangle$ remains in its instantaneous eigenstate during the evolution driven by the slowly varying system Hamiltonian $H(t)$.
    Therefore, adiabatic protocols are normally designed according to the associated quantitative adiabatic condition, which can be generally written as
    \begin{equation}\label{eq:adia}
        \left| \frac{\langle \Psi_m(t)| \dot{H}(t)|\Psi_n(t) \rangle}{(E_n(t) - E_m(t))^2}\right| \ll 1
    \end{equation}
    where $|\Psi_m(t)\rangle$ ($|\Psi_n(t)\rangle$) is the eigenstate of $H(t)$ with corresponding eigenenergy $E_m(t)$ ($E_m(t)$). $\dot{H}(t)$ is the derivative of Hamiltonian with respect to time $t$. 
    This adiabatic condition indicates that the implementation of the adiabatic control is challenging for quantum systems with short coherence times and weak anharmonicity such as superconsucting quantum circuits~\cite{PhysRevA.76.042319, PhysRevLett.111.080502}.
	Fortunately, it is found that Eq. \eqref{eq:adia} may not govern every adiabatic processes~\cite{PhysRevLett.93.160408, sarandy2004consistency, PhysRevLett.95.110407, PhysRevLett.101.060403, PhysRevLett.102.220401, PhysRevLett.104.120401, li2014quantitative}.
	This stimulated extensive works to design an adiabatic evolution that follows new adiabatic conditions instead of the restriction of Eq.~(\ref{eq:adia}) \cite{wu_adiabatic_2007, PhysRevA.76.024304, PhysRevA.80.012106, PhysRevLett.98.150402, lidar2009adiabatic, PhysRevA.81.032308, PhysRevA.85.062111, PhysRevA.88.012114, PhysRevA.93.052107, PhysRevLett.115.133601, russomanno_floquet_2017, PRXQuantum.2.030302}.
    Recently, 
	a sufficient and necessary adiabatic condition (SNAC) has been proposed \cite{PhysRevA.93.052107},
	which can replace the traditional condition of Eq. \eqref{eq:adia}.
	SNAC may be realized by modulating the dynamic phases \cite{xu2019breaking}.
    Its theoretical foundation is very different from that of previous studies, which rely on energy gaps \cite{PhysRevA.76.024304, PhysRevA.80.012106, PhysRevLett.98.150402, lidar2009adiabatic, PhysRevA.81.032308, PhysRevA.85.062111, PhysRevA.88.012114} or Floquet resonances \cite{PhysRevLett.115.133601, russomanno_floquet_2017}.
	It is also different from recent work on the adiabatic modulation theorem which is based on slow modulations of rapidly varying fields \cite{PRXQuantum.2.030302}.
	SNAC suggests a possibility to increase the speed and hence the fidelity of adiabatic control. 

    In this article, we demonstrate quantum adiabatic transfer between one-qubit and two-qubit quantum states by designing appropriate unitary control fields and periodically evolving the state along geodesics~\cite{PhysRevA.93.052107} in superconducting quantum circuits.
	These adiabatic evolutions fulfill the sufficient and necessary adiabatic condition, supporting the modification of the traditional adiabatic condition.
	Our experimental data show that this protocol can achieve high-fidelity quantum state transfer with a short operation time.
    Moreover, we demonstrate the control robustness of this protocol by applying noise through qubit control line.
	Our protocol provides an alternation to the traditional adiabatic control method for quantum state manipulation in superconducting circuits.

\section{Protocol description}
\label{sec:protocol}

	\begin{figure}
		\begin{minipage}[b]{0.5\textwidth}
			\centering
			\includegraphics[width=8.5cm]{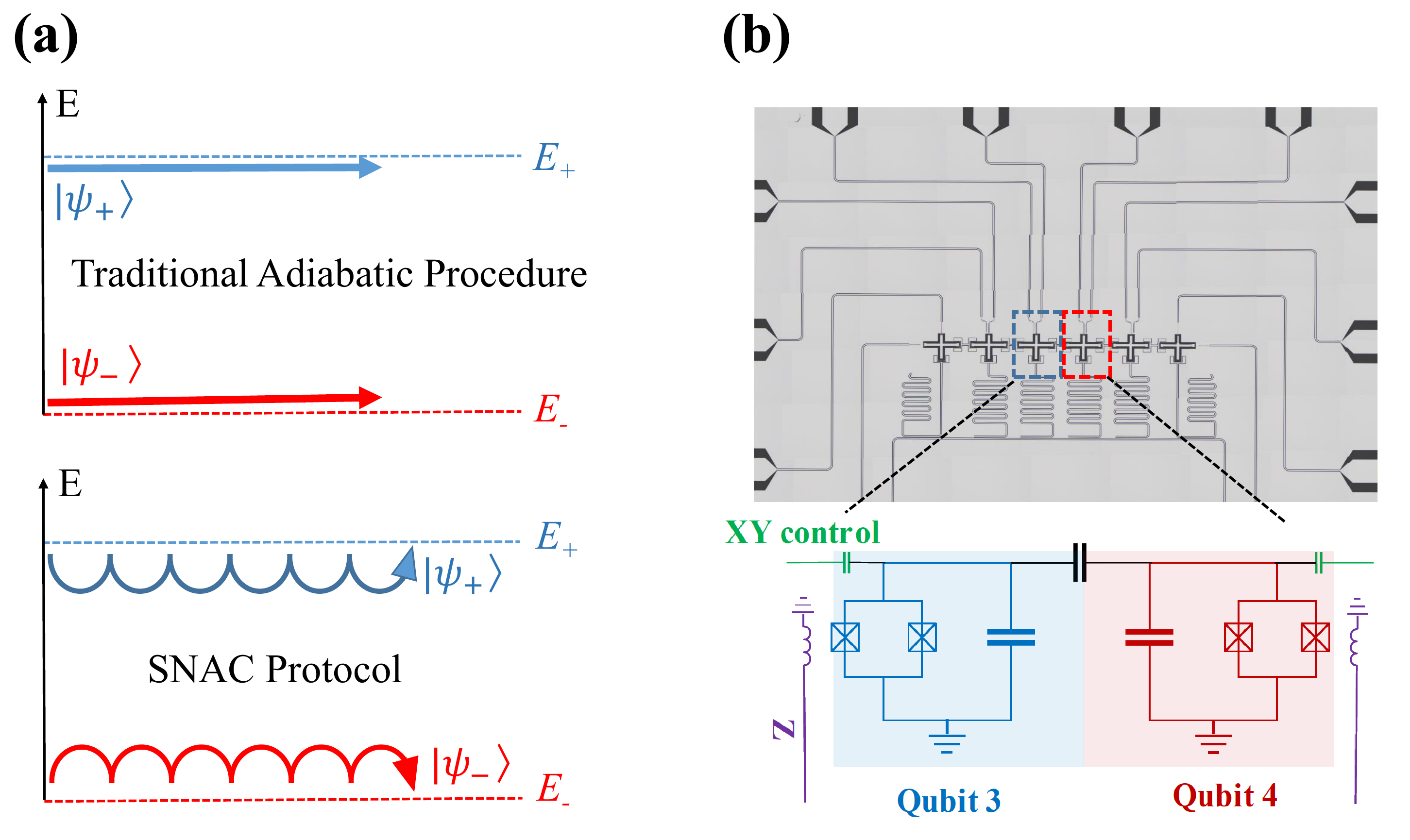}
		\end{minipage}
		\caption{(Color online)	\textbf{(a)}$\,$Schematic of the quantum adiabatic process.
			In the traditional adiabatic evolution process, the quantum state evolves alongside the eigenstate.
			However, in the protocol based on SNAC, the quantum state evolves around the eigenstate periodically. 
			\textbf{(b)}$\,$Full image of the quantum chip we used in experiment. 
			The center two of six transmons are numbered as Qubit 3 ($Q_3$) and Qubit 4 ($Q_4$), highlighted by the red and blue dashed boxes with enlarged equivalent circuits at the bottom. 
			\label{fig:FIG1}
		}
	\end{figure}

	Instead of Eq. \eqref{eq:adia}, a different adiabatic approximation condition is proposed by Wang and Plenio \cite{PhysRevA.93.052107}. 
	They decompose the system dynamics driven by a Hamiltonian into an ideal quantum evolution in the adiabatic limit and a diabatic propagator that includes all the diabatic errors. An adiabatic path can be built
	when the dynamic phase factors at different path points add destructively and tend to vanishing, which is written as 
	\begin{equation}\label{eq:unitaryControlAdiaCond}
	\left| \int_{0}^{\lambda} e^{i\chi_{n,m}(\lambda ')}d\lambda ' \right| \rightarrow 0,
	\end{equation}
	where $\chi_{n,m}(\lambda)\equiv\chi_n(\lambda)-\chi_m(\lambda)$ is the difference of the accumulated dynamic phases on the eigenstate, $\lambda$ is the configuration parameter evolving the system.
	In our experiments, $\lambda$ corresponds to an angle and is tuned in time.
	If our system evolves from $t=0$ to $t=T$, we can divide the whole process into $N$ segment.
	Then we design the evolution parameter as 
	\begin{equation}\label{eq:jumpingPulse}
	\lambda(t) = [\lambda(T)-\lambda(0)]\left(\frac{2Int(Nt/T)+1}{2N}\right) + \lambda(0),
	\end{equation}
	where $Int\equiv[$ $]$ is integral function. Therefore, $\lambda(t)$ is a discontinuous function of time. We can introduce a jumping ratio
	$\gamma \equiv \frac{\delta \lambda}{(\lambda(T) -\lambda(0))/N}$ to illustrate the discontinuity in control field, where	
	$\delta \lambda \in [0, \frac{\lambda_T -\lambda_0}{N}]$ defines the minimum step in the parameter space for each segment, resulting a corresponding discontinuity on the evolution path.
	In general, a finite $\delta \lambda$ makes the Hamiltonian jump forward in parameter space, evolving the quantum state periodically, as illustrated in the bottom panel of Fig.~\ref{fig:FIG1}(a). When $\delta \lambda=0$, it looks like that the path collapses to that of the traditional adiabatic evolution. However, the traditional adiabatic condition Eq.~\eqref{eq:adia} could still be violated. On the other hand, when $N \to \infty$, $\delta \lambda \to 0$ and the discontinuity also vanishes. Now intuitively the SNAC protocol is equivalent to the traditional protocol since both Eq. \eqref{eq:adia} and Eq. \eqref{eq:unitaryControlAdiaCond} are satisfied simultaneously, regardless of the jumping ratio value.

	Eq. \eqref{eq:unitaryControlAdiaCond} shows that the differences of dynamic phases $\chi_{n,m}$ are more fundamental than the energy gaps in adiabatic evolution because the energy does not explicitly appear in the condition.
	If we choose $\chi_{n,m} = \pi$ in each segment to satisfy Eq.~\eqref{eq:unitaryControlAdiaCond}, a perfect adiabatic evolution can be realized, even though the traditional adiabatic condition  Eq.~(\ref{eq:adia}) is violated.
	Fig.~\ref{fig:FIG1}(a) illustrates the differences of evolution path between two protocols.
	Therefore, it is plausible to design a shorter evolution time $T$ while obtaining high-fidelity state transfer by designing suitable unitary control fields to satisfy Eq. \eqref{eq:unitaryControlAdiaCond}.

\section{Experiments and Results}

	The superconducting quantum circuit we used is composed of six tunable and grounded transmons \cite{PhysRevA.76.042319, PhysRevLett.111.080502}, as shown in Fig.~\ref{fig:FIG1}(b).
	More details of the chip layout and control wiring from room temperature to 10 mK can be found in the supplementary note of our previous works \cite{zheng_optimal_2022}.
	In our experiment, we select two of transmons numbered as $Q_3$ and $Q_4$ to demonstrate quantum state transfer. Their parameters are summarized in Table~\ref{supptable:device_params}.

	\begin{table}[htbp]
		\begin{ruledtabular} 
			\begin{threeparttable}  
				\caption{Device parameters.} 
				\label{supptable:device_params}
				\begin{tabular}{c|ccc}
					\hspace*{2cm} & $Q_3$ & $Q_4$ \\\hline  \\[-0.3cm]
					$\omega_{\mathrm{q}}/2\pi$\tnote{a}~ (GHz) & 4.9559 & 5.1866 (5.8428) \\ 
					$\alpha_{\mathrm{q}}/2\pi$\tnote{b}~ (GHz) & -0.286 & -0.268\\ 
					$\omega_{\mathrm{r}}/2\pi$\tnote{c}~ (GHz) & 6.9099 & 7.0655\\ 
					$\kappa_{\mathrm{r}}/2\pi$\tnote{d}~ (MHz) & 4.189  & 1.750\\ 
					$\chi_{\mathrm{r}}/2\pi$\tnote{e}~ (MHz) & 0.044  & 0.168 (0.259)\\ 
					$T_1$\tnote{f}~ (\SI{}{\micro\second}) & 13.4  & 22.1 (26.4)\\ 
					$T_2^*$\tnote{g}~ (\SI{}{\micro\second})& 10.8  & -- (45.4) \\ 
				\end{tabular}
				\footnotesize
				\begin{tablenotes}
					\item[a] Transition frequency $|0\rangle \rightarrow |1\rangle$ at the working spot (sweet spot).
					\item[b] anharmonicity.
					\item[c] Readout resonator frequency.
					\item[d] Readout resonator linewidth. The difference between linewidths $\kappa_r$ is mainly caused by the intrinsic qualities of the resonators.
					\item[e] Effective dispersive shift at the working spot (sweet spot) for the $|0\rangle \rightarrow |1\rangle$ transition due to the interaction with the readout cavity mode.
					\item[f] Energy decay time ($T_1$) measured at the working spot (sweet spot).
					\item[g] Ramsey decay time ($T_2^*$) measured at the working spot (sweet spot).
				\end{tablenotes}
			\end{threeparttable}  
		\end{ruledtabular} 
	\end{table}

	\begin{figure}
		\begin{minipage}[b]{0.5\textwidth}
			\centering
			\includegraphics[width=8cm]{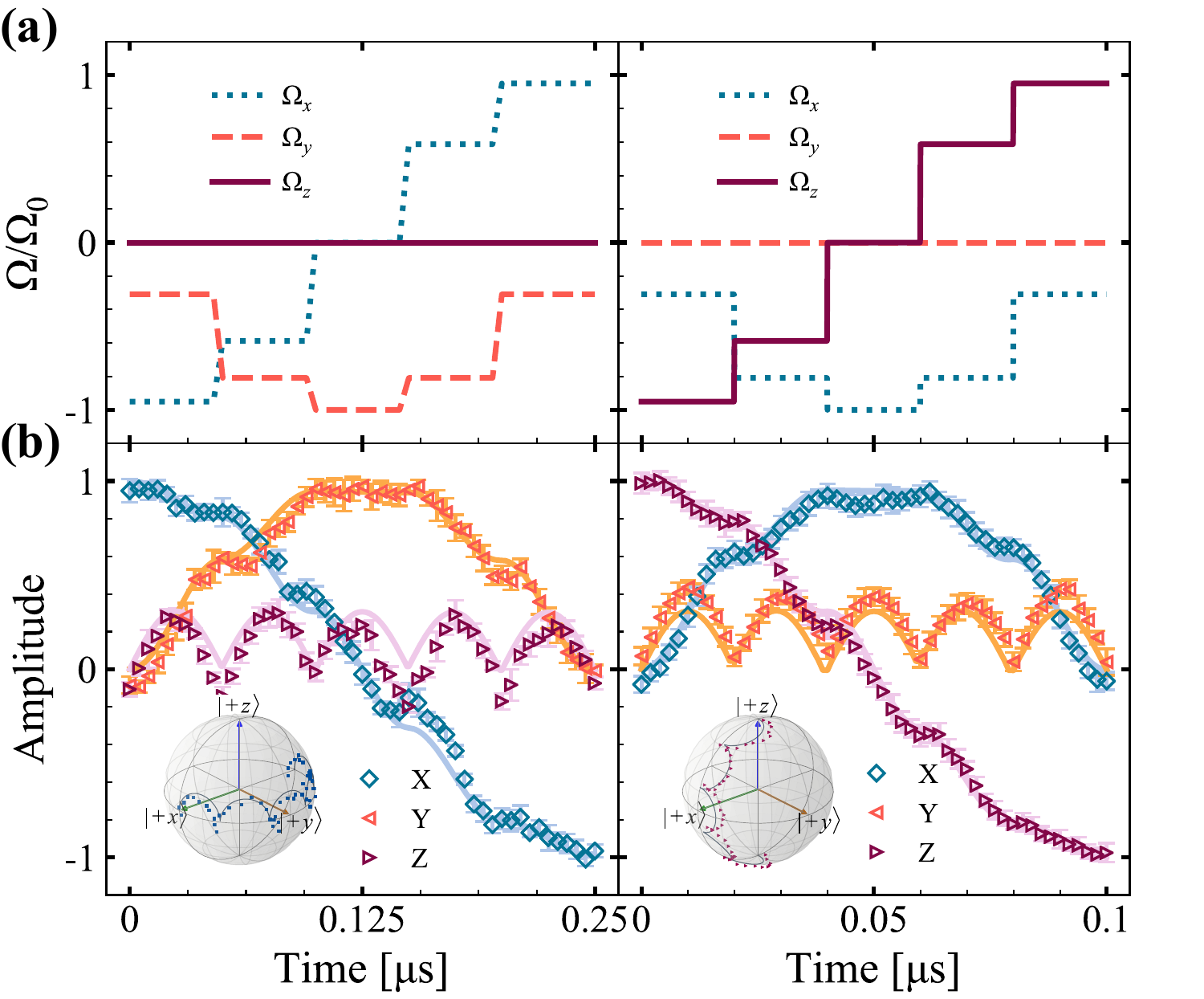}
		\end{minipage}
		\caption{(Color online) State transfers in one-qubit system.
			\textbf{(a)}$\,$ The amplitude of three components (corresponding to the projection along three Pauli matrices) of microwave field $MW(t)$.
			The evolution process in left (right) panel is designed along latitude (longitude).
			\textbf{(b)}$\,$State evolution during the transfer process.
			The initial state in left (right) panel is prepared on $|+x\rangle$ ($|+z\rangle$) by a unitary rotation $R_y(\frac{\pi}{2})$ ($I$), corresponding to the instantaneous eigenstate $|\psi_{-}(0)\rangle$ with parameters $\theta=\pi/2$ and $\phi=-\pi$ ($\theta=-\pi$ and $\phi=0$) at $t=0$.
			The Bloch spheres illustrate the corresponding evolution trajectories of state transfers.
			\label{fig:FIG2}
		}
	\end{figure}
		
    We first demonstrate state transfer using the SNAC protocol on the transmon qubit labeled as $Q_4$, which has the dephasing time $T_2^{\star} = 45.4\,\SI{}{\micro\second}$ measured by Ramsey fringes experiment and the relaxation time $T_1 = 26.4\, \SI{}{\micro\second}$.
	In our experiment, by introducing a driving microwave field $MW(t) = \Omega(t) \cos(\omega_d(t) t+\phi(t))$, 
	the transmon Hamiltonian is written as $H(t) = \omega_{4} a^{\dagger}a + \frac{\alpha_{4}}{2} a^{\dagger}a^{\dagger}aa + MW(t)(a^{\dagger}+a)$, 
	where  $a^{\dagger}$ ($a$) is the creation (annihilation) operator,
	$\phi(t)$ is the phase of the microwave field, and
	$\theta(t) = \arctan{\frac{\Omega(t)}{\Delta(t)}}$ is the mixing angle.
	The amplitude ($\rm{Max}\left(|\Omega(t)|\right) /2\pi= 25$ MHz) is far smaller than the anharmonicity of the transmon ($|\alpha_4|/2\pi = 268$ MHz).
	By using rotation wave approximation, the Hamiltonian $H(t)$ can be parameterized as
	\begin{equation}\label{twoLevelsH}
		H_1(t) = \frac{\Omega_0}{2}(t)\left(
			\begin{array}{cc}
				\cos{\theta(t)} & \sin{\theta(t)}e^{-i\phi(t)} \\
				\sin{\theta(t)}e^{i\phi(t)} & -\cos{\theta(t)} \\
			\end{array}
		\right),
	\end{equation} 
	where 
	$\Omega_0(t) = \sqrt{\Delta(t)^2 + \Omega(t)^2}$ is the energy gap of $H_1(t)$,
	and $\Delta(t) = \omega_4 - \omega_d(t)$
	is the detuning between the driving frequency $\omega_d(t)$ and the transition frequency $\omega_4$ of the two lowest levels in the transmon.
	The eigenenergies of $H_1(t)$ are $\pm \frac{1}{2}\Omega_0(t)$ with the eigenstates
	$|\psi_+(t) \rangle = \cos\frac{\theta(t)}{2} |0\rangle + \sin\frac{\theta(t)}{2} e^{i \phi(t)} |1\rangle$ and 
	$|\psi_-(t) \rangle = \sin\frac{\theta(t)}{2} |0\rangle - \cos\frac{\theta(t)}{2} e^{i \phi(t)} |1\rangle$.
	By designing appropriate parameter ranges for $\phi(t)$ and $\theta(t)$ from $-\pi$ to $0$, we demonstrate high-fidelity state transfers based on the evolution of instantaneous eigenstates $|\psi_-(t)\rangle$.

    In order to fulfill Eq. \eqref{eq:unitaryControlAdiaCond}, we design the evolution path by using Eq. \eqref{eq:jumpingPulse} with different parameters. In our experiment, we use $N=5,\,\lambda(0)=-\pi$, and $\lambda(T)=0$ and set the evolution time as $T = 5\pi/\Omega_0$.
	The components of the applied waveform $\vec{\Omega}(t) = \mathrm{Tr}[H_1(t)\vec{\sigma}]/2$ are obtained and shown in Fig. \ref{fig:FIG2}(a).
	Here, $\vec{\sigma} = {\sigma_x, \sigma_y, \sigma_z}$ are the Pauli matrices.
	To disclose the trajectory along latitude (longitude) in the entire evolution, we apply the protocol with state initialized at $|+x\rangle$ ($|+z\rangle$).
	Finally, the corresponding waveform for quantum state tomography (QST) is applied to measure the evolution trajectory.
	
	As shown in Fig. \ref{fig:FIG2}(b), the state evolution along latitude (longitude) by set $\Omega_0(t)/2\pi = 10\, \rm{MHz}$ ($25\, \rm{MHz}$) is in good agreement with the simulation results. 
	The trajectories on Bloch sphere are plotted in the inset.
	Here, we point out that the error bars indicate standard deviation in this article.
	At the end of the evolution, the fidelities of the target states $\rho_{ideal} = |-x\rangle \langle -x|$ (latitude) and $\rho_{ideal} = |-z\rangle \langle -z|$ (longitude) are $0.983 \pm 0.019$ and $0.989 \pm 0.026$ respectively, which are benchmarked by
	\begin{equation}\label{eq:fidelity}
		\mathcal{F} = \mathrm{Tr}\sqrt{\sqrt{\rho_{ideal}} \rho_{exp} \sqrt{\rho_{ideal}}},
	\end{equation}
	where $\rho_{exp}$ is the experimental final state.
	The corresponding fidelities in the simulation are $0.992$ and $0.996$, respectively.
	Since the fidelities of preparing  $|+x\rangle$ and $|+z\rangle$ are $0.985 \pm 0.028$ and $0.995 \pm 0.028$, the limitations of the state transfer along latitude are mainly caused by state preparation and measurement errors, while the errors when evolving along longitude are caused by measurement errors due to both the imperfections of QST and the errors of quantum state readout. In order to further increase the fidelity, one may use optimal pulses sequence to eliminate state leakage. 
	It is worth noting that although there are tiny deformations of trajectories during evolution processing,
	the highest quality state transfer can still be obtained, indicating the control robustness of this protocol.

	\begin{figure}
		\begin{minipage}[b]{0.5\textwidth}
			\centering
			\includegraphics[width=8.0cm]{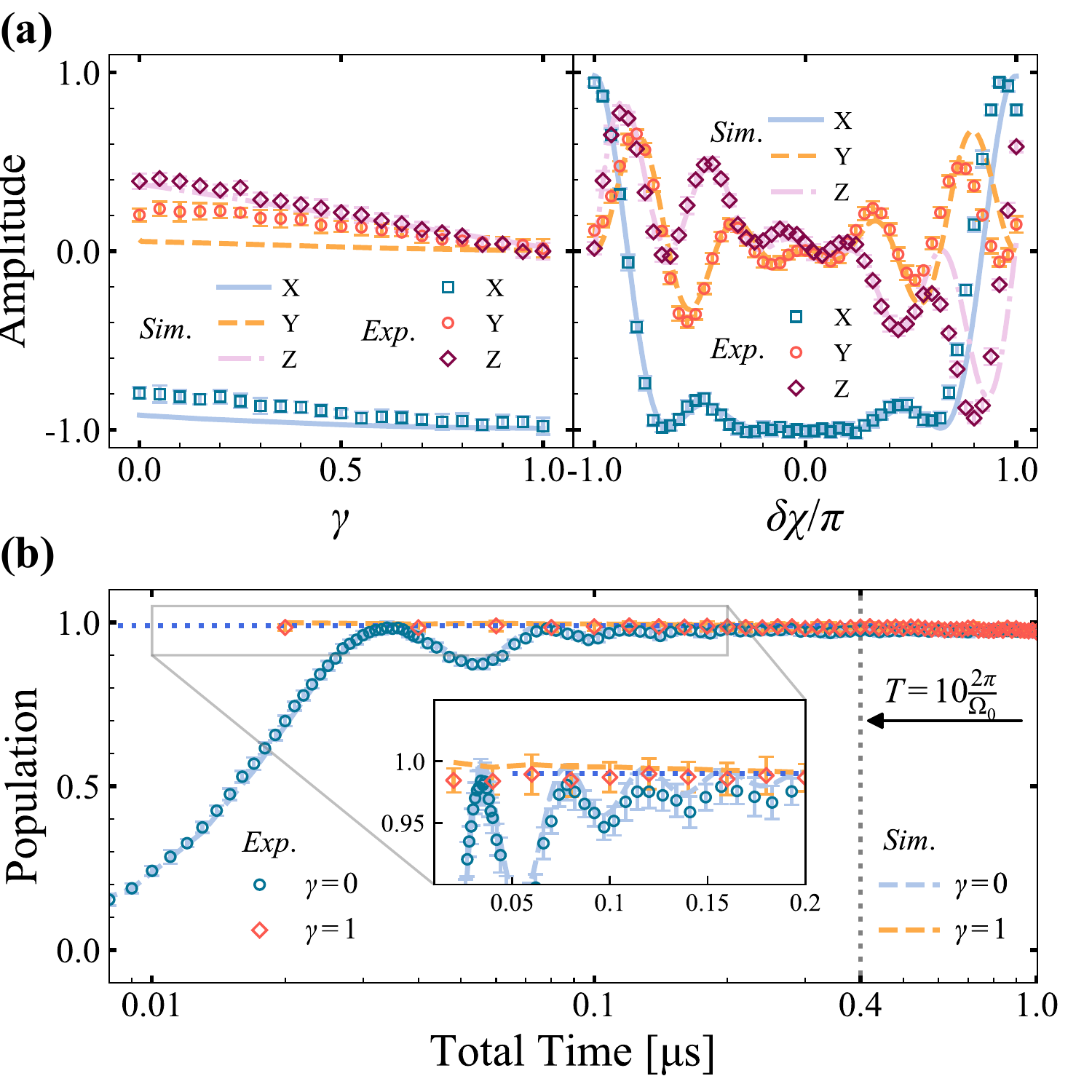}
		\end{minipage}
		\caption{(Color online) Performance of the state transfer.
			\textbf{(a)}$\,$Transferred state $\rho_{exp}$ vs. the jumping ratio $\gamma$ (the left panel) and the dynamic phase $\chi$ (the right panel).
			The evolution process is along latitude by set $\Omega_0/2\pi = 10\, \rm{MHz}$ and $T = 250$ ns.
			\textbf{(b)}$\,$Population of the state transfer vs. the total time $T$.
			The evolution process is along longitude by set $\Omega_0/2\pi = 25\, \rm{MHz}$.
			The dashed line marks the 10 times of the period of Rabi oscillations. This is approximately the traditional limit of adiabatic evolution since using these parameters we calculate that the left-hand side of Eq. \eqref{eq:adia} is about 0.05.
			\label{fig:FIG3}
		}
	\end{figure} 

    To explore the theoretical boundary of this protocol, we investigate the evolution process along latitude for different jumping ratio $\gamma$.
	As shown in the left panel of Fig. \ref{fig:FIG3}(a), the instantaneous state $|\psi_{-}(T)\rangle$ approaches the target state $|-x\rangle$ when the jumping ratio $\gamma$ varies from $\gamma=0$ to $\gamma=1$.
	In all these procedures, $\Omega_0(t)/2\pi = 10\, \rm{MHz}$ and the total evolution time $T = 250\, \rm{ns}$ are used,
	so the left-hand side of Eq. \eqref{eq:adia} is about $0.2$, indicating that these procedures dissatisfy the restriction of Eq. \eqref{eq:adia}.
	When $\gamma=0$, the evolution trajectory is the same as that of the traditional adiabatic transfer. The fidelity is about 90$\%$. The highest fidelity is achieved at $\gamma=1$, meaning that the adiabatic speed limit can be broken by jumping along the geodesic if the dynamic phase can be canceled carefully\cite{xu2019breaking}.

	Furthermore, we demonstrate how the dynamic phase $\chi\equiv\chi_{n,m}$ (we omit $n,m$ here after) affects state transfer in the SNAC protocol. As illustrated in the right panel of Fig. \ref{fig:FIG3}(a), 
	the larger the phase offset $|\delta \chi| = |\chi - \pi|$, the more deviation from the target state $|-x\rangle$, indicating that a zero phase is essential to satisfying the adiabatic approximation. Therefore, a satisfactory adiabatic process can be obtained by designing appropriate phase $\chi$ and evolution time $T$.
    If we compare the transfer fidelity for $\gamma=0$ and $\gamma=1$ as a function of total evolution time $T$, the SNAC protocol can be about 4 times faster than the traditional method, as shown in Fig. \ref{fig:FIG3}(b). The acceleration of the adiabatic evolution process is of interest to fundamental physics as well as applications in quantum control and quantum annealing. 
	We point out that two transitions overlap with each other at the time about $T=400$ ns (marked as the dashed line), where traditional adiabatic condition is satisfied. If we calculate the left-hand side of Eq. \eqref{eq:adia} using experimental parameters we obtain 0.05, fulfilling the traditional limit of adiabatic control.

	\begin{figure}
		\begin{minipage}[b]{0.5\textwidth}
			\centering
			\includegraphics[width=7.5cm]{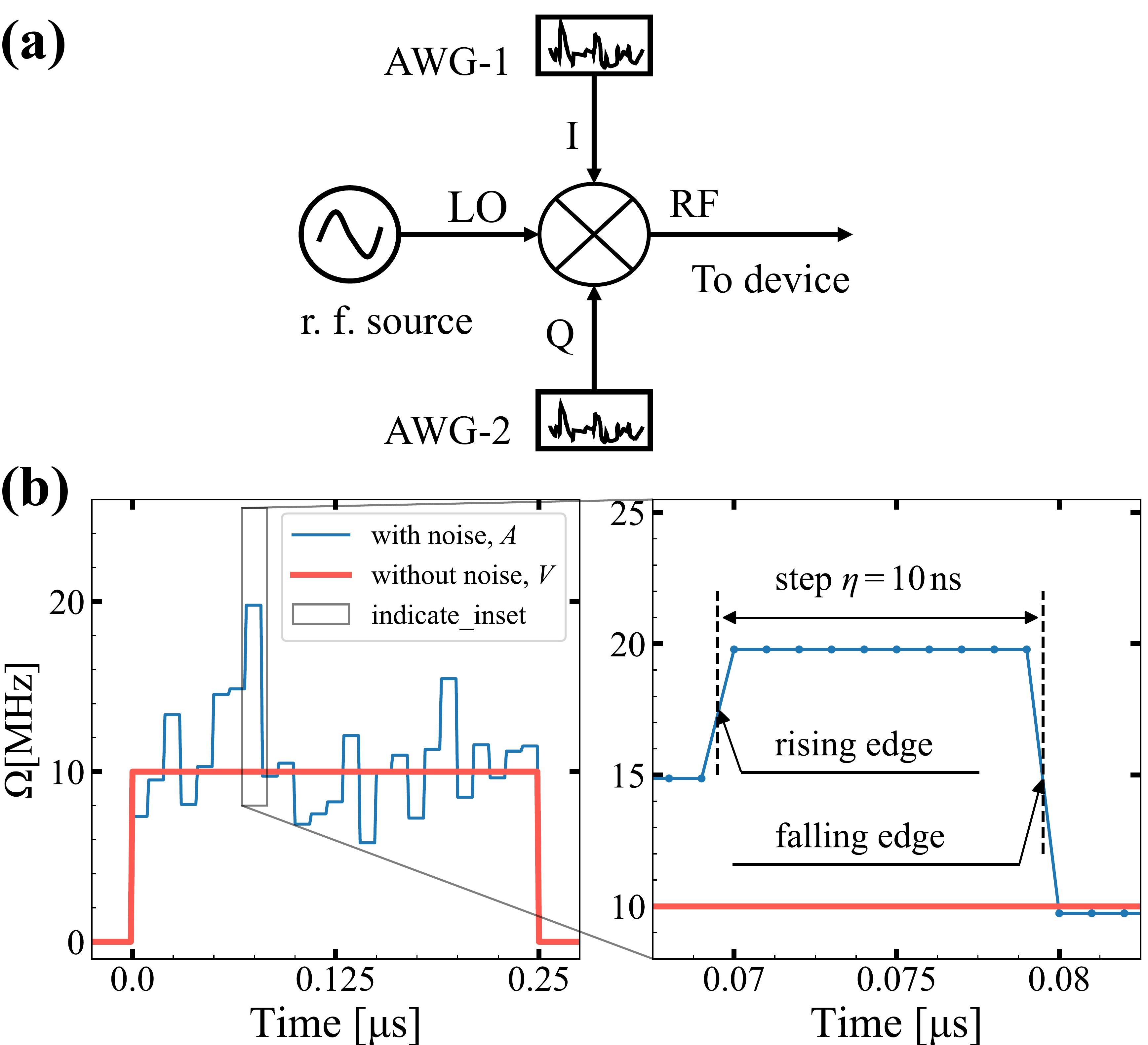}
			\includegraphics[width=8.0cm]{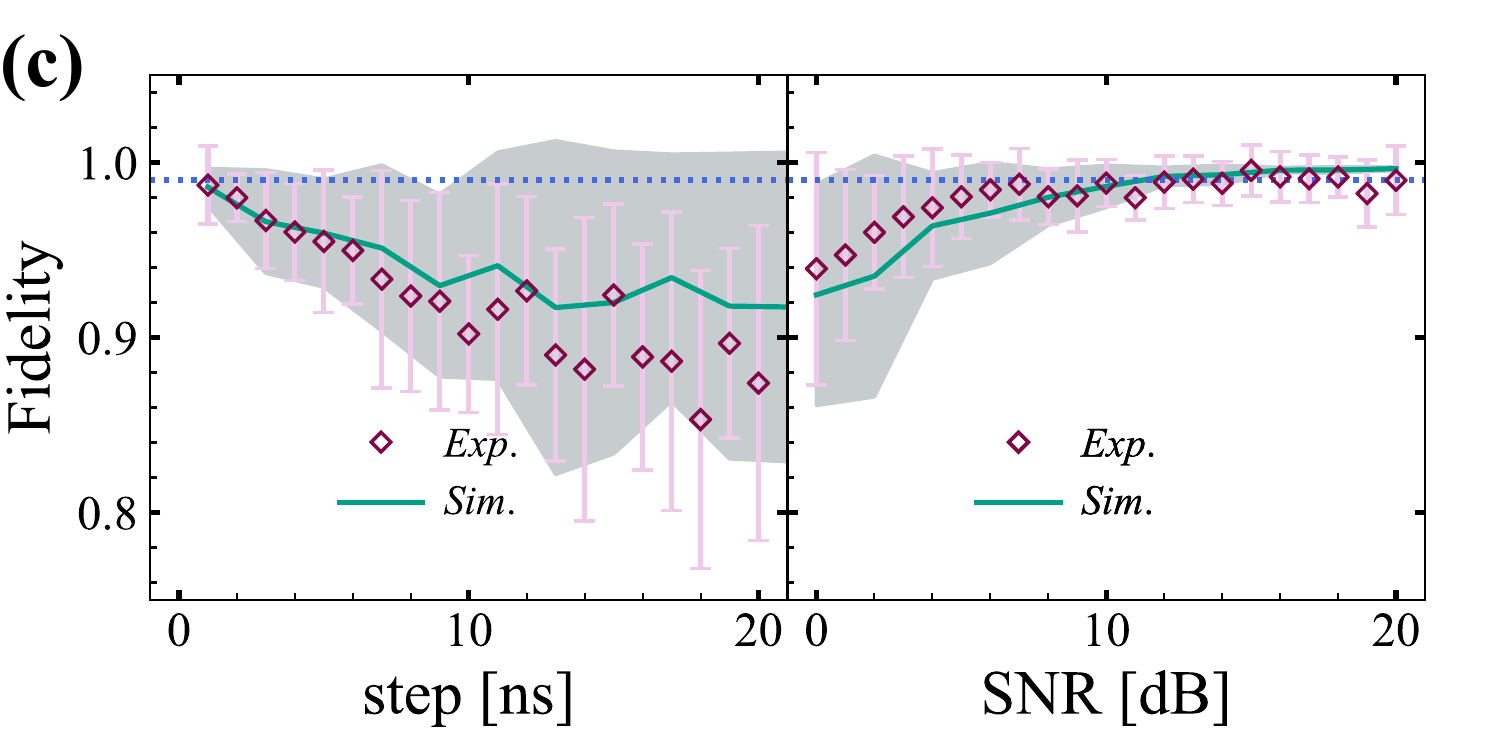}
		\end{minipage}
		\caption{(Color online) Generation of Gaussian noise and its effect on the transfer fidelity.
			\textbf{(a)}$\,$Schematic of the circuit to synthesize the waveform with Gaussian noise. 
			A sinusoidal continuous wave signal generated by r.f. source sends to the local oscillator port in an IQ mixer,
			of which the intermediate frequency ports receive two quadrature signals with the digital Gaussian noise generated by an AWG (Keysight model M3202A, 14 bits, 1 GSa/s).
			Then the synthesized waveform is output from the radio frequency port and sent to the qubit control line.
			\textbf{(b)}$\,$Waveform $V(t)$ with/without Gaussian noise with the parameters of noise: $\rm{SNR}=10\, \rm{dB}$, and $\eta = 10\, \rm{ns}$.
			\textbf{(c)}$\,$(left panel) Fidelity of the transfer $\rho_{exp}$ as a function of the step with $\rm{SNR}=10\, \rm{dB}$. (right panel) Fidelity of the transfer vs. noise power with $\eta = 1\, \rm{ns}$ .
			The points (green lines) are the experimental (simulation) results, while the gray zones mark the simulation standard deviations.
			\label{supp_fig:SM_WGN_demo}
			}
	\end{figure}

	Next, we demonstrate the robustness of this protocol against experimental perturbations. 
	Gaussian noise \cite{kafadar1986gaussian}, which mimics the disordered noise from the environment of experimental setup, is applied by mixing the signals generated by arbitrary waveform generator (AWG), as shown in Fig. \ref{supp_fig:SM_WGN_demo}(a).
	The AWG model we used is Keysight M3202A (14 bits, 1 GSa/s), and the r.f. source model is R$\&$S SGS-100A.
	In our experimental setup, it is difficult to directly realize Gaussian noise using analog hardware.
	Therefore, the digital synthesis method \cite{kafadar1986gaussian} is introduced to add Gaussian noise to the ideal digital signals, which can be converted to analog signals by AWG.
	Without loss of generality, we assume that the digital signals contain signal voltages $V = [V_0, V_1, \dots, V_{M-1}]$ 
	and noise voltages $A = [A_0, A_1, \dots, A_{M-1}]$. Then the signal-to-noise ratio (SNR) can be defined as
	\begin{equation}\label{SNRDefine}
		\mathrm{SNR} = 10 \lg{\frac{\mathrm{P}_V}{\mathrm{P}_A}} = 10 \lg{\sum_{\mu}\frac{V_{\mu}^2}{A_{\mu}^2}},
	\end{equation}
	where 
	the signal (noise) power is labeled as $\mathrm{P}_V$ ($\mathrm{P}_A$),
	and $V_{\mu}$ ($A_{\mu}$) is the signal (noise) amplitude of the $\mu$-th digital point, which corresponds to the AWG voltage.
	We further use the \textit{randn} function in Python to generate the random number $\aleph = [\aleph_0, \aleph_1, \dots, \aleph_{L-1}]$, then the noise voltage can be expressed as
	\begin{equation}\label{eq:NoiseVoltsCalc}
		A_{\mu} = V_{\mu}(1 + \aleph_l \sqrt{\frac{10\sum_{\mu}{V_{\mu}}}{\mathrm{SNR}}}); \quad \mathrm{for}\ \nu \in [0, \eta], \quad \mu = l\eta+\nu,
	\end{equation}
	with $\mu \in [0, M-1]$, $l\in [0, L-1]$, and $L = \frac{M}{\eta \times \mathrm{sr}}$.
	Here, all indices $\{M,\, L,\, l,\, \mu,\, \nu\}$ take integers only, 
	the noise voltage is set as a constant value for each noise pulse with holding time $\eta$,
	and $\mathrm{sr} = 1\,\mathrm{GSa/s}$ denotes the AWG sampling rate.
	As shown in the left panel of Fig. \ref{supp_fig:SM_WGN_demo}(b), the blue line is the noise pulse with noise parameters: $\rm{SNR} = 10$ dB, $\eta = 10$ ns, while the red line is the perfect signal pulse. The detail shape of a typical noise pulse is illustrated in the right panel in Fig. \ref{supp_fig:SM_WGN_demo}(b). The rising and falling edges are 1 ns and the holding time $\eta$ is 10 ns. 
	We measured the transfer fidelity as a function of $\eta$, and the results are shown in the left panel of Fig. \ref{supp_fig:SM_WGN_demo}(c). It is observed that the holding time of the noise affects the transfer fidelity dramatically. The reason is that the noise amplitude increases with the $\eta$ when the total noise power is fixed.
	Then we keep at the optimized condition $\eta=1$ ns, and measure the transfer fidelity as a function of the noise power.  
	As shown in the right panel of Fig. \ref{supp_fig:SM_WGN_demo}(c), the experimental results maintain high fidelity in a wide range of SNR (larger than 10 dB), which is close to $0.99$ as denoted by the dotted blue line.
	The excellent noise suppression indicates that this protocol has the potential to obtain high-fidelity state transfers in the Gaussian noise environment.

	\begin{figure}
		\begin{minipage}[b]{0.5\textwidth}
			\centering
			\includegraphics[width=7.5cm]{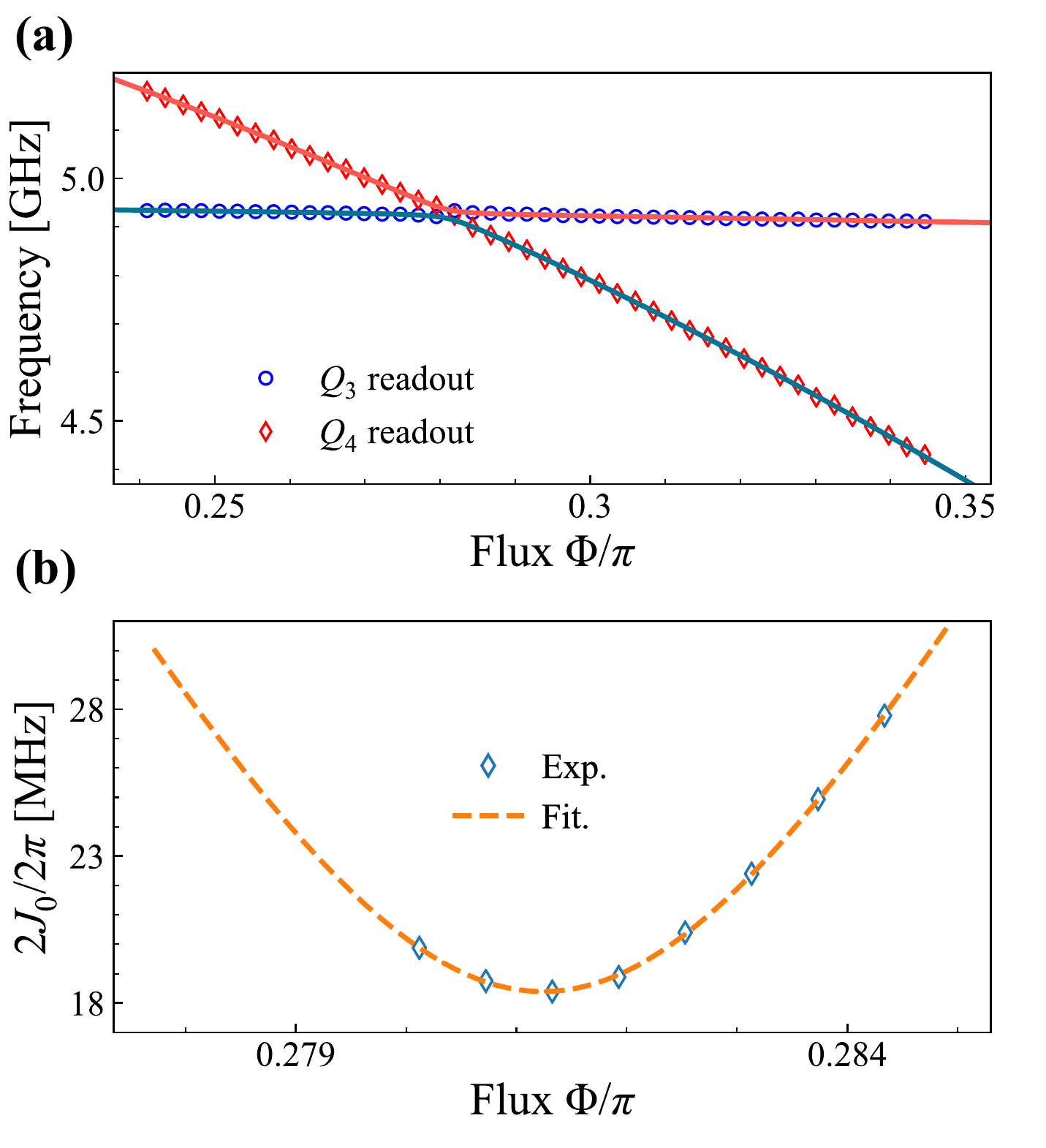}
		\end{minipage}
		\caption{(Color online) Calibration of the detuning. \textbf{(a)}$\,$The spectra of $Q_3$ and $Q_4$ versus flux bias of $Q_4$. 
			The symbols are experimental results extracted by spectroscopy measurement, while the lines are the eigenvalues of the two-qubit Hamiltonian in subspace $\{|01\rangle, |10\rangle\}$.
			\textbf{(b)}$\,$Measurement of the effective coupling strength.
			The experimental results (symbols) are best fitted by $2J_{0}(t) = \sqrt{4J^2 + \delta^2(t)}$ with $J/2\pi = 9.2\, \rm{MHz}$.
			\label{supp_fig:SM_tqcali}
			}
	\end{figure}
			
	\begin{figure}
		\begin{minipage}[b]{0.5\textwidth}
			\centering
			\includegraphics[width=8.0cm]{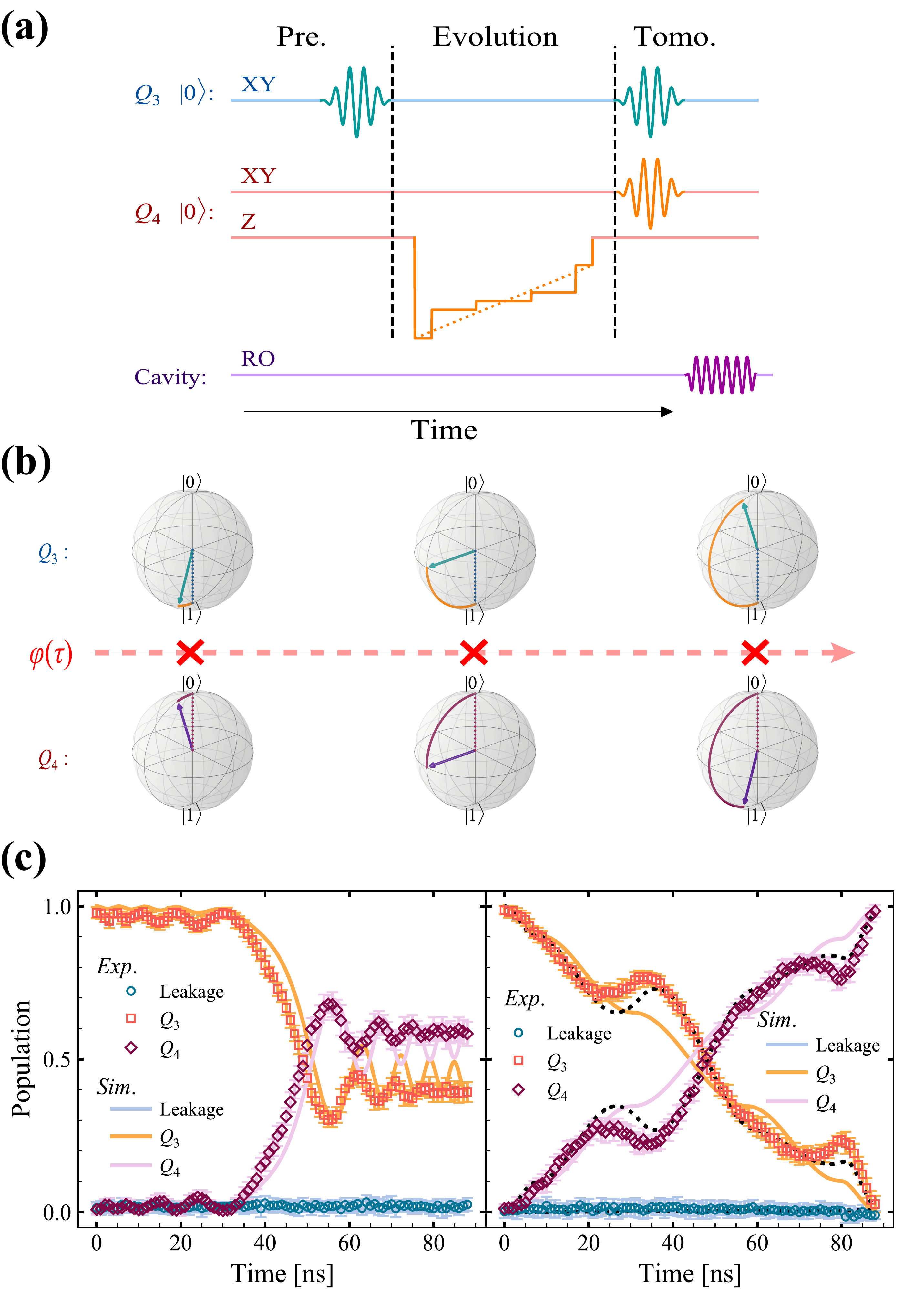}
		\end{minipage}
		\caption{(Color online) State transfer in two-qubit system.
			\textbf{(a)}$\,$Diagram of the experimental time profile.
			These include: preparing the initial states of qubits, evolving the system state with the designing flux pulse to obtain the state transfer from $Q_3$ to $Q_4$, and measuring the evolution trajectories by QST.
			\textbf{(b)}$\,$Illustration of state transfers with $\varphi = \pi/10,\, 5\pi/10$, and $9\pi/10$.
			\textbf{(c)}$\,$State transfers based on LZT and SNAC with evolution time $T=88\, \rm{ns}$.
			The population of the excited state in $Q_4$ based on LZT extracted from experimental (simulation) results is $0.583\pm0.040$ ($0.588$), while that based on SNAC is $0.984\pm0.020$ ($0.986$).
			\label{fig:FIG4}
		}
	\end{figure}

	Finally, we demonstrate state transfers between two transmons, where $Q_3$ ($Q_4$) acts as a fix (tunable) qubit, whose decoherence times are $T_1 = 13.4\,\SI{}{\micro\second}\,(22.1\,\SI{}{\micro\second})$ and $T_2^{\star} = 10.8\,\SI{}{\micro\second}\, (4.5\,\SI{}{\micro\second})$.
	Since the transmons are capacitively coupled with coupling strength $J/2\pi=9.2\, \rm{MHz}$, the Hamiltonian can be written as 
	$H = \omega_3 a^{\dagger} a + \frac{\alpha_{3}}{2} a^{\dagger} a^{\dagger} a a + \omega_4 b^{\dagger} b + \frac{\alpha_{4}}{2} b^{\dagger} b^{\dagger} b b + J(a^{\dagger} + a)(b^{\dagger} + b)$,
	where 
	$b^{\dagger}$ ($b$) is the creation (annihilation) operator.
    By using the rotating wave approximation, the system Hamiltonian in subspace $\{|00\rangle,\, |01\rangle,\, |10\rangle,\, |11\rangle\}$ can be simplified as 
	\begin{equation}\label{eq:twoQubitsH}
	H_2(t) = J_0(t)
		\left(
			\begin{array}{cccc}
				0 & 0 & 0 & 0\\
				0 & \cos{\varphi(t)} & \sin{\varphi(t)} & 0\\
				0 & \sin{\varphi(t)} & -\cos{\varphi(t)} & 0\\
				0 & 0 & 0 & 0\\
			\end{array}
		\right)
	\end{equation}
	where in the interest subspace $\{|01\rangle, |10\rangle\}$,
	the eigenenergies are $E_{\pm}(t) = \pm J_{0}(t)$ with $J_{0}(t) = \sqrt{\delta^2(t)/4 + J^2}$,
	the eigenstates are
	$|E_{+}(t)\rangle = \cos{\frac{\varphi(t)}{2}}|01\rangle + \sin{\frac{\varphi(t)}{2}}|10\rangle$ and 
	$|E_{-}(t)\rangle = \sin{\frac{\varphi(t)}{2}}|01\rangle - \cos{\frac{\varphi(t)}{2}}|10\rangle$,
	while $\delta(t) = \omega_{4}(t) - \omega_{3}$ is the detuning between $Q_3$ and $Q_4$,
	and $\varphi(t) = \arctan{\frac{2J}{\delta(t)}}$ denotes the parameterized phase, which can be obtained through calibrating the detuning $\delta(t)$.

	We calibrated the coupling between two qubits first. As shown in Fig. \ref{supp_fig:SM_tqcali}(a), by changing the bias flux of the tunable qubit $Q_4$ we can observe an energy level avoided crossing due to the coupling between two qubits. Then we carefully measured the detuning $\delta(t)$ and coulping strength $J$ using vacuum Rabi oscillations in subspace $\{|01\rangle, |10\rangle\}$, as shown in Fig. \ref{supp_fig:SM_tqcali}(b).

	In our experiment, the SNAC protocol is implemented in two-qubit system by modulating the parameter $\varphi(t)$ to satisfy Eq. \eqref{eq:unitaryControlAdiaCond}.
	The measurement sequence is shown in Fig. \ref{fig:FIG4}(a),
	which is composed of three sections: preparation for the initial state $|\psi_0\rangle$, evolution based on the flux pulse designed according to Eq. \eqref{eq:jumpingPulse} and Eq. \eqref{eq:unitaryControlAdiaCond}, and measurement with QST to reconstruct the final states.
	At $t=0$, the qubit $Q_3$ ($Q_4$) is biased at $\omega_3/2\pi = 4.9559\, \rm{GHz}$ ($\omega_4/2\pi = 5.1866\, \mathrm{GHz}$) with the prepared excited (ground) state by applying a unitary rotation $R (\pi)$ ($I$).
	The detuning $\delta(0)/2\pi = 230.7\, \mathrm{MHz}$ is far larger than the coupling strength, so the unwanted transfers during preparation and measurement can be neglected.
	Therefore, the state transfer between $Q_3$ and $Q_4$ with the SNAC protocol can be realized by designing $\varphi (t)$ under conditions: $\gamma=1$, $N=5$, $\lambda(0) = 0$, and $\lambda(T) = \pi$. The profile of XY control pulse and Z flux bias of the qubits are shown in Fig. \ref{fig:FIG4}(a).
	Here, parameter $\varphi (t)$ reveals the evolution path of state transfers in parameter space, as shown by the evolution diagram in Fig. \ref{fig:FIG4}(b).
	We compare our protocol with the one based on the traditional LZT \cite{landau1932, zener1932non}, which is fundamental to the dynamics of quantum systems including adiabatic evolution.
    With an identical flux bias range and evolution time ($T=88\, \rm{ns}$), the SNAC protocol has significantly better performance than the LZT scheme as shown in Fig. \ref{fig:FIG4}(c). 
	The corresponding populations of the final state $\rho_{Q_4} = \rm{Tr}_{Q_3}(\rho_{Q_3 Q_4})$ are $0.583\pm 0.040$ and $0.984\pm 0.020$, respectively.
	In the right panel of Fig. \ref{fig:FIG4}(c), the deviations between experimental results (symbols) and simulation results (orange solid lines) are mainly due to the distortions of the flux pulse, which lead to accumulated phase errors $\delta \chi$ hence violating the condition in Eq. \eqref{eq:unitaryControlAdiaCond}.
	To verify this argument we do numerical simulation by adding phase errors $\delta \chi = [0,\, 0.28\pi,\, 0.15\pi,\, 0.1\pi,\, 0]$ to each segment thus obtaining a mock deformed flux pulse.
	Here, the phase errors are obtained by changing the spectrum of $Q_4$ while fixing the evolution time.
	As shown by the black dotted lines, the trajectories with phase errors are better matching with the experiment results. Alternatively, we can add corresponding phase offsets by calibrating the actual dynamic phase $\chi$ in each segment to cancel the errors.

\section{Discussion}\label{sec:Discussion}
	\begin{figure}
		\begin{minipage}[b]{0.5\textwidth}
			\centering
			\includegraphics[width=7.5cm]{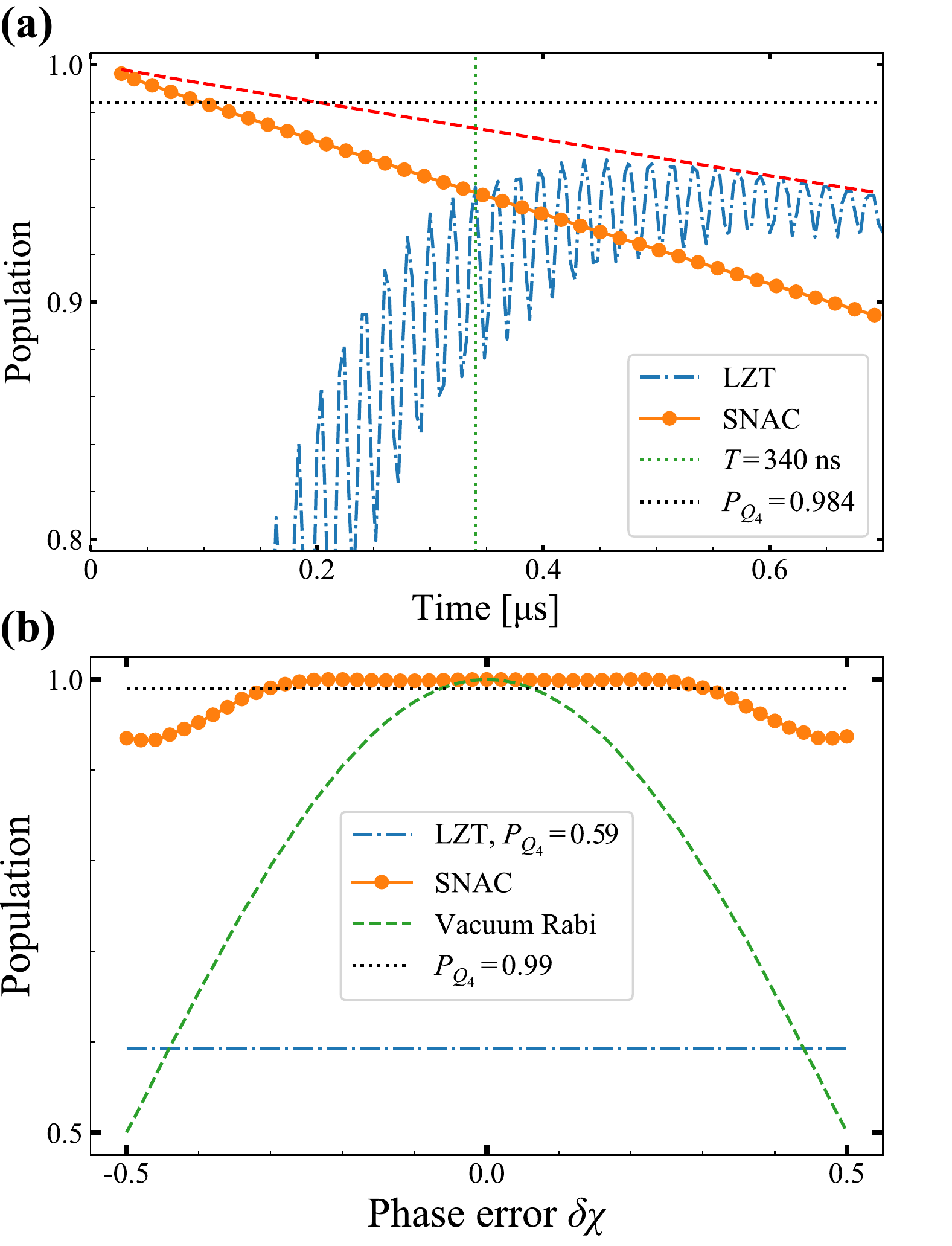}
		\end{minipage}
		\caption{(Color online)
			\textbf{(a)}$\,$Numerical simulation of transferred population with respect to evolution time for different schemes.
			Due to the energy relaxation and dephasing, the population of the excited state of $Q_4$ approaches maximum $0.948$ at the evolution time $T > 340\, \rm{ns}$ for LZT.
			The transferred population based on SNAC decreases exponentially with increasing the evolution time. 
			The horizontal black dotted line marks the population of $0.984$ as a reference.
			The vertical green dash-dotted line marks the evolution time at $340\, \rm{ns}$.
			The red dashed line shows the upper boundary of the transferred population limited by the energy relaxation time.
			\textbf{(b)}$\,$Demonstration of the robustness of SNAC scheme. 
			The orange symbols are the simulated transferred population with respect to the dynamical phase deviation $\delta \chi$ for SNAC scheme.
			The simulation parameters are: $\gamma = 1$, $N=5$, $\lambda_0=0$, and $\lambda_T=\pi$.
			The green dashed line shows the calculated transfer population using iSWAP operation based on resonant vacuum Rabi oscillations. 
			The blue dash-dotted line indicates the results $P_{Q_4} = 0.59$ based on LZT and the black dotted line marks a reference with values $0.99$.
			\label{supp_fig:StateTranVsTime}
			}
	\end{figure}
	To further analysis the performance of SNAC scheme compared with LZT and dynamic protocols, we calculated population transfer vs. time and phase deviation.
	We choose the decoherence times of a two-qubit system at working spots as listed in Table~\ref{supptable:device_params} and the dephasing time of the tunable qubit as $\bar{T}_2=4.5\,\SI{}{\micro\second}$, which is the average dephasing time of $Q_4$ and also close to a practical value for the state-of-art quantum chips. The calculated transferred populations vs. time are shown in Fig. \ref{supp_fig:StateTranVsTime}(a). 
	For LZT, due to the limitation of the adiabatic condition in Eq. \eqref{eq:adia}, the population increases with transfer time and approaches maximum $0.95$ at about $400$ ns.
	Then it will decrease due to the effect of energy relaxation.
	The transferred population using our SNAC protocol can achieve $0.995$ at $50$ ns, indicating that the evolution constrained by Eq. \eqref{eq:unitaryControlAdiaCond} can be much faster than that of LZT.
	The population exponentially decreases with the evolution time due to the decoherence.
	Since the SNAC protocol is realized by modulating the dynamic phase, besides of the energy relaxation, the dephasing also affects the transferred population, hence resulting in that
	the asymptotic value at long time is lower than that of LZT. However, in our experiments the population of the excited state of $Q_4$ for LZT can never achieve $0.96$ due to the energy relaxation.
	Therefore, the SNAC protocol will have much better performance for the system with relatively short decoherence times. 
	In addition, SNAC provides a faster tool to realize state transfer because it would take about $\sim 340\, \rm{ns}$ to reach the same fidelity using LZT. 
	
	The robustness of the population transfer is represented by the transferred population vs. the deviation of the control parameter, e.g., phase $\chi$.
	We calculated the transferred populations with respect to the phase error $\delta \chi$ without considering decoherence, shown in Fig. \ref{supp_fig:StateTranVsTime}(b).
	Since the protocol based on LZT has nothing to do with the phase $\chi$, its result is always about $0.59$ (the blue dashed-dotted line).
	On the contrary, the protocol based on iSWAP gate, which is realized by resonant vacuum Rabi oscillations, is sensitive to the phase error $\delta \chi$ because it is completely a non-adiabatic process.
	The results (the orange symbols) based on SNAC show advantage in resisting the phase error.
	Therefore, the SNAC protocol possesses the advantage of speed compared with usual LZT scheme while insensitive to the fluctuation of the control parameter due to its adiabatic nature. Considering quantum state transfer is ubiquitous in quantum computation and simulation, the protocol based on SNAC provides a useful tool for quantum information process with superconducting qubits. 

\section{Conclusions} \label{sec:conclusions}
    In summary, we demonstrated adiabatic quantum state transfers in one and two superconducting qubits by canceling the dynamic phase.
	The principle we apply is very different from that of traditional adiabatic evolution.
	Our results prove that one can mitigate the constraint of adiabatic condition in multi-qubit systems and accelerate the evolution process while keeping it immune to the Gaussian noise of the system.
    The speed and robustness of state transfers make our protocol a promising control tool in quantum computation and quantum simulation.

\section*{Acknowledgements}\label{sec:acknowledgements}
	This work was partly supported by the Key R\&D Program of
	Guangdong Province (Grant No. 2018B030326001), NSFC (Grant No. 61521001, No. 12074179, No. 11890704, and No. U21A20436), and NSF of Jiangsu Province (Grant No. BE2021015-1).



	
\end{document}